\documentclass[prb,aps,twocolumn,tightenlines,showpacs,floatfix,amsmath,amssymb]{revtex4}

\usepackage{graphicx}
\usepackage{dcolumn}
\usepackage{bm}
\usepackage{colordvi}
\begin{document}

\title{Control of spin coherence in semiconductor double quantum
  dots}
\author{Y. Y. Wang}
\affiliation{Hefei National Laboratory for Physical Sciences at
  Microscale and Department of Physics, University of Science and Technology of China, Hefei,
  Anhui, 230026, China}
\author{M. W. Wu}
\thanks{Author to whom correspondence should be addressed}%
\email{mwwu@ustc.edu.cn.}
\affiliation{Hefei National Laboratory for Physical Sciences at
  Microscale and Department of Physics, University of
Science and Technology of China, Hefei,
  Anhui, 230026, China}

\date{\today}
\begin{abstract}
We propose a scheme to manipulate the spin coherence in
vertically coupled GaAs double quantum dots. Up to {\em
 ten} orders of magnitude variation of the spin relaxation and
 {\em two} orders of magnitude variation of the spin dephasing can
 be achieved  by a small gate voltage applied vertically on the double
 dot. Specially, large variation of spin relaxation still exists at
  $0$\ K. In the calculation, the equation-of-motion approach is
  applied to obtain the electron decoherence time and all the relevant
spin decoherence mechanisms, such as the spin-orbit coupling together
  with the electron--bulk-phonon scattering, the direct spin-phonon
  coupling due to the phonon-induced strain, the hyperfine interaction
  and the second-order process of electron-phonon scattering combined
  with the hyperfine interaction, are included.
  The condition to obtain the large variations of spin
  coherence is also addressed.

\end{abstract}
\maketitle

\section{Introduction}

The fast development of spintronics aims at making devices based on
the electron spin. Semiconductor quantum dots (QDs) are one of the promising
candidates for the implementation of quantum
computations\cite{Awschalom,Hans,Loss,Taylor0} because
of the relative long spin coherence time, which has been
proved both theoretically\cite{Jiang2,Semenov} and
experimentally.\cite{Hanson,Amasha,Petta}
Among different kinds of QDs, double quantum dot (DQD)
system  attracted much more attention recently as there is an
additional coupling
between two QDs in both vertical\cite{Pi,Ono,Austing} and
parallel\cite{Petta,Koppens,Laird,jaro} DQDs,
 which can be controlled conveniently by a small
gate voltage. Therefore spin devices
based on DQDs can be designed with more flexibility.  So far many
elements of the spintronic device,
such as quantum logical gates,\cite{Mason,Hanson2} spin
filters\cite{Cota,Mireles}
and spin pumps\cite{Cota} were
proposed and/or demonstrated based on DQD system.
Specially, in our previous work, a way to control spin
relaxation time ($T_1$) induced by the spin-orbit coupling (SOC)
together with the
electron--bulk-phonon (BP) scattering in DQDs by a small gate voltage
was proposed.\cite{Wang} However, according to our latest study,\cite{Jiang2}
 the spin relaxation can be controlled
by other mechanisms also, if calculated correctly.
In the present paper, we include all the spin decoherence mechanisms
following our latest study in the single QD system,
 and apply the equation-of-motion approach to study the spin
decoherence in DQD system. By this approach, not only the
 spin relaxation time but also the spin dephasing time ($T_{2}$) can
 be obtained.  We show that both the spin
relaxation and the spin dephasing can be manipulated by a small gate
voltage in DQD system.
Especially, the large variation of spin relaxation still
exists even at $0$\ K. The DQD system studied here can be
realized easily using the present available technology.

We organize the paper as following. In Sec.\ II we set up the model and
briefly introduce different spin decoherence mechanisms. The
equation-of-motion approach is also explained simply. Then in Sec.\ III
we present our numerical results. We first show
how the eigen energies and
eigen wave functions vary with the bias field in Sec.\ IIIA. Then
the electric field dependences
of spin relaxation and spin dephasing are shown in Sec.\ IIIB and
IIIC, respectively. We conclude in Sec.\ IV.

\section{Model and  Method}

We consider a single electron spin in two vertically coupled QDs with
a bias voltage $V_{d}$ and an external magnetic field
${\mathbf B}$
applied along the growth direction ($z$-axis). Each QD is confined by
a parabolic potential $V_{c}({\bf r})
=\frac{1}{2} m^{\ast} \omega_{0}^{2} {\bf r}^{2}$
(therefore the effective dot diameter
$d_{0}=\sqrt{\hbar\pi/m^{\ast} \omega_{0}}$) in the
$x$-$y$ plane  in a quantum well of width $d$.
The confining potential $V_{z}(z)$ along the $z$-direction reads
\begin{equation}
V_{z}(z)=
\begin{cases} eEz+\frac{1}{2}eV_{d}, & \frac{1}{2}a<|z|<\frac{1}{2}a+d\\
eEz+\frac{1}{2}eV_{d} +V_{0}, &|z|\leqslant\frac{1}{2}a \\
\infty, & \mbox{otherwise}
\end{cases}\ ,
\end{equation}
in which $V_{0}$ represents the barrier height between the
two dots, $a$ stands for the
inter-dot distance and $E=V_{d}/(a+2d)$ denotes the
electric field due to the bias voltage $V_{d}$.
Then the electron Hamiltonian reads
$H_{e}= \mathbf{P}^{2}/(2m^\ast) +
  V_c({\mathbf r}) + V_{z}(z)+ H_Z + H_{SO}$,
where $m^{\ast}$ is the
electron effective mass and $\mathbf{P}=-i \hbar \mathbf{\nabla} +
\frac{e}{c} \bf{A}$ with
${\mathbf A}=(B/2)(-y,x,0)$ being the vector potential.
 $H_Z=\frac{1}{2}g\mu_B B \sigma_{z}$ is the Zeeman
energy with $g$, $\mu_{B}$ and $\mbox{\boldmath
  $\sigma$\unboldmath}$ being the $g$-factor
of electron, Bohr magneton and Pauli matrix respectively.
$H_{so}$ is the Hamiltonian of the SOC.
In GaAs, when the quantum well width and the gate voltage
along the growth direction are small, the Rashba SOC\cite{Yu} is
unimportant.\cite{flat} Therefore, only the
Dresselhaus term\cite{Dresselhaus} contributes to $H_{SO}$. When the
quantum well width is smaller than the QD diameter, the
dominant term in the Dresselhaus SOC reads
$H_{so}= \frac{1}{\hbar}\sum_{\lambda}\gamma_{\lambda}^{\ast}
  (-P_{x}\sigma_{x}+P_{y}\sigma_{y})$,
with $\gamma_{\lambda}^{\ast} = \gamma_{0} \langle P_{z}^{2}\rangle
 /\hbar^{2}$. $\gamma_{0}$ denotes the Dresselhaus
 coefficient, $\lambda$ is the quantum number of $z$-direction and $\langle
P_{z}^{2} \rangle_\lambda \equiv -\hbar^2 \int
\psi_{z,\lambda}^\ast(z)\partial^2/\partial
z^2\psi_{z,\lambda}(z)dz$, where $\psi_{z,\lambda}$
($\lambda=1,2,3\cdots$) is the eigen wave function of the electron
along the $z$-direction.\cite{Wang}
The electron eigen
energy and wave function in the $x$-$y$--plane can be obtained by the exact
 diagonalization approach.\cite{Cheng}

The interactions between the electron and the lattice
lead to the electron spin decoherence. These interactions
contain two parts, one
is the hyperfine interaction between the electron and the nuclei, the
other is the electron-phonon interaction which is further composed of the
electron-BP interaction $H_{ep}$, the direct spin-phonon coupling due
to the phonon induced strain $H_{strain}$ and the phonon-induced
$g$-factor fluctuation.
We  briefly summarize these spin decoherence mechanisms and
the detailed expressions can be found in Ref.\ \onlinecite{Jiang2}:
(i) The SOC together with the electron-BP scattering $H_{ep}$. As the
SOC mixes different spins, the electron-BP can induce spin relaxation
and spin dephasing. (ii) Direct
spin-phonon coupling due to the phonon-induced stain
$H_{strain}$.\cite{Dyakonov} Because this mechanism mixes different spins
and also is related to the electron-phonon interaction,
it can induce spin decoherence {\em alone}.
 (iii) The hyperfine interaction $H_{eI}$.\cite{Abragam}
 It is noted that the hyperfine
interaction alone only induces $T_{2}$ since it only changes the
electron spin, but not the electron energy.
 (iv) The second-order process of hyperfine
interaction combined with the electron-BP interaction $V_{eI-ph}^{(3)}=  | \ell_{2}
\rangle[\sum_{m\not=\ell_1}( \langle
\ell_{2} | H_{ep} |m \rangle \langle m | {H_{eI}({\mathbf r})} |\ell_{1}
\rangle)/ (\varepsilon_{\ell_{1}} - \varepsilon_{m}) +
\sum_{m\not=\ell_2}( \langle
\ell_{2} |{H_{eI} ({\mathbf r})} |m \rangle \langle m |H_{ep}|\ell_{1}
\rangle)/(\varepsilon_{\ell_{2}} -\varepsilon_{m} )]\langle \ell_{1}|$, where
$|\ell_{i}\rangle$ ($i=1,2,3\cdots$) and $\varepsilon_{\ell_{i}}$ are
the eigen state and eigen energy of $H_{e}$, respectively.
Although the hyperfine interaction can not induce $T_{1}$ alone,
it is noted that the second-order process,
 combined with the electron-BP interaction,  can induce
both $T_{1}$ and $T_{2}$.
The other mechanisms, including the first-order process of hyperfine
interaction combined with the electron-BP scattering\cite{Abalmassov} and the
$g$-factor fluctuation\cite{Roth} have been proved to
be negligible.\cite{Jiang2}

Due to the SOC, all the states are impure spin states
with different expectation values of the
spin. For finite temperature, the electron is distributed over many
states and therefore one has to
average over all the involved
processes  to obtain the total
spin relaxation time. The Fermi-Golden-Rule approach calculates the
spin relaxation from the initial state to the final one whose
majority spin polarizations are opposite. However, the average method is
inadequate when many impure states are
included.\cite{Jiang2} Also
the Fermi-Golden-Rule approach can not be used to calculate
 the spin dephasing time.
Therefore, in this paper we adopt the
equation-of-motion approach for many-level
system with Born approximation developed in
Ref.\ \onlinecite{Jiang2}.
 When the spin
dephasing induced by the hyperfine interaction is considered, as the slow
relaxation of nuclear bath compared to the electron, the kinetics is
non-Markovian. Moreover,
because of the Born approximation, this equation-of-motion approach
can only be applied for strong magnetic field ($\ge 3.5$\
T).\cite{Coish} Therefore, the
pair-correlation method\cite{Yao} is further adopted to calculate the
hyperfine interaction induced $T_{2}$ for small magnetic field.
What should be emphasized is that the SOC is always
included, no matter which mechanism is considered.
It has been shown that its effect to spin decoherence can not
be neglected.\cite{Jiang2}

 \begin{figure}[bth]
  \centering
    \includegraphics[height=8cm,width=7.5cm]{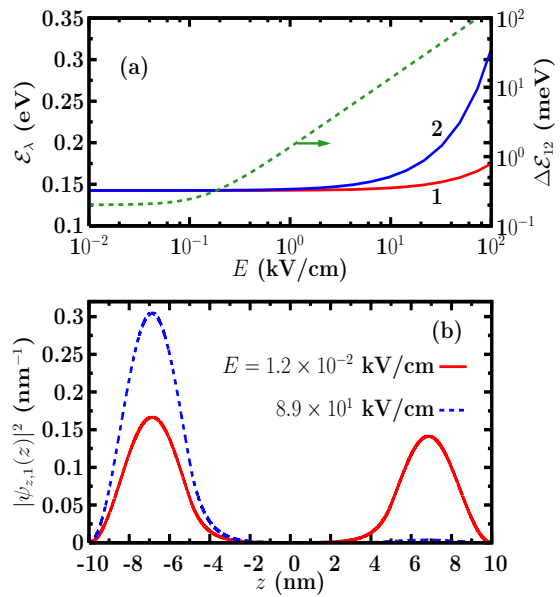}
\caption{(Color on line) (a) The Lowest two eigen energies along the
  $z$-axis ${\cal E}_{\lambda}$ ($\lambda=1,2$) and the energy
  difference $\Delta {\cal E}_{12}={\cal E}_{2}-{\cal E}_{1}$ {\em
    vs.} the bias field $E$. Note the scale of $\Delta {\cal E}_{12}$ is on the
  right side of the figure. (b) Square of the absolute value of the ground
  state wave function along the $z$-axis at two typical bias fields. In the
  calculation, the well width $d=5$\ nm, the inter-dot distance $a=10$\ nm and
  the barrier height $V_{0}=0.4$\ eV. }
\label{fig1}
\end{figure}

\section{Numerical Results}

Following the method addressed above, we perform the numerical
calculation in a typical vertically
coupled GaAs DQD with barrier height
$V_{0}=0.4$\ eV, inter-dot distance $a=10$\ nm and
well width $d=5$\ nm.
The GaAs material parameters and the parameters
related to different mechanisms are the same with those in Ref.\
\onlinecite{Jiang2}.

\subsection{Electric field dependence of eigen energy and
  eigen wave function  along $z$-axis}

The eigen energy and eigen wave function along the $z$-direction are
obtained numerically. In Fig.\ \ref{fig1}(a), the lowest two eigen
energies ${\cal E}_{1}$ and ${\cal E}_{2}$ along the $z$-axis and their
difference $\Delta {\cal E}_{12}=
{\cal E}_{2}-{\cal E}_{1}$ are plotted as functions of the electric
field $E$. It can be seen that the energy difference $\Delta {\cal E}_{12}$
increases quickly with
the electric field $E$ when $E$ is larger than $0.1$\ kV/cm.
The eigen wave function of the ground state along the $z$-axis also has
large variation with $E$. It can be seen clearly in Fig.\
  \ref{fig1}(b) that when $E$ is very small
($1.2 \times 10^{-2}$\ kV/cm), the wave function of the ground state
locates at the two wells almost equally. However, when $E$ is large
enough ($89$\ kV/cm), the wave function of the ground
state locates mostly at the quantum well with lower potential.
The physics of such bias-voltage-induced quick change of eigen energy
and eigen wave function can be understood as what follows.
 Because of the large barrier height $V_{0}$ and/or
large inter-dot distance $a$, the two quantum dots are nearly
independent and the eigen wave function along the
$z$-axis of the lowest subband spreads equally over the two QDs
when the source-drain
voltage is very small. Therefore at this time the energy difference
between the lowest two energy levels along the $z$-axis $\Delta {\cal E}_{12}$
is very small. However, with the
increase of the source-drain voltage, electron can tunnel through the barrier
 and the wave function is almost located at one dot with
lower potential and
therefore $\Delta {\cal E}_{12}$ increases quickly with $E$.

\begin{figure}[bth]
  \centering
    \includegraphics[height=4.2cm,width=5.1cm]{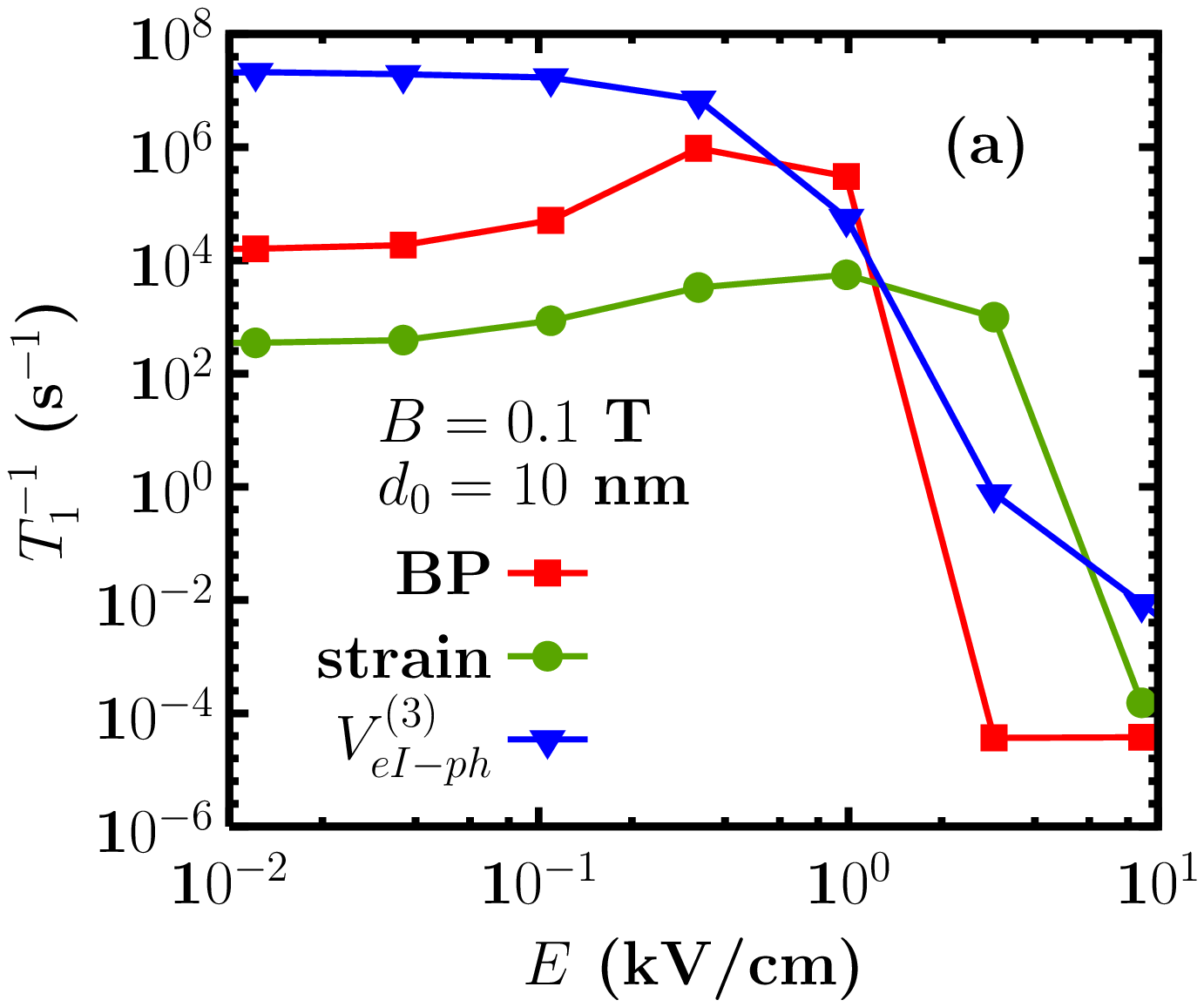}
    \includegraphics[height=4.2cm,width=5.1cm]{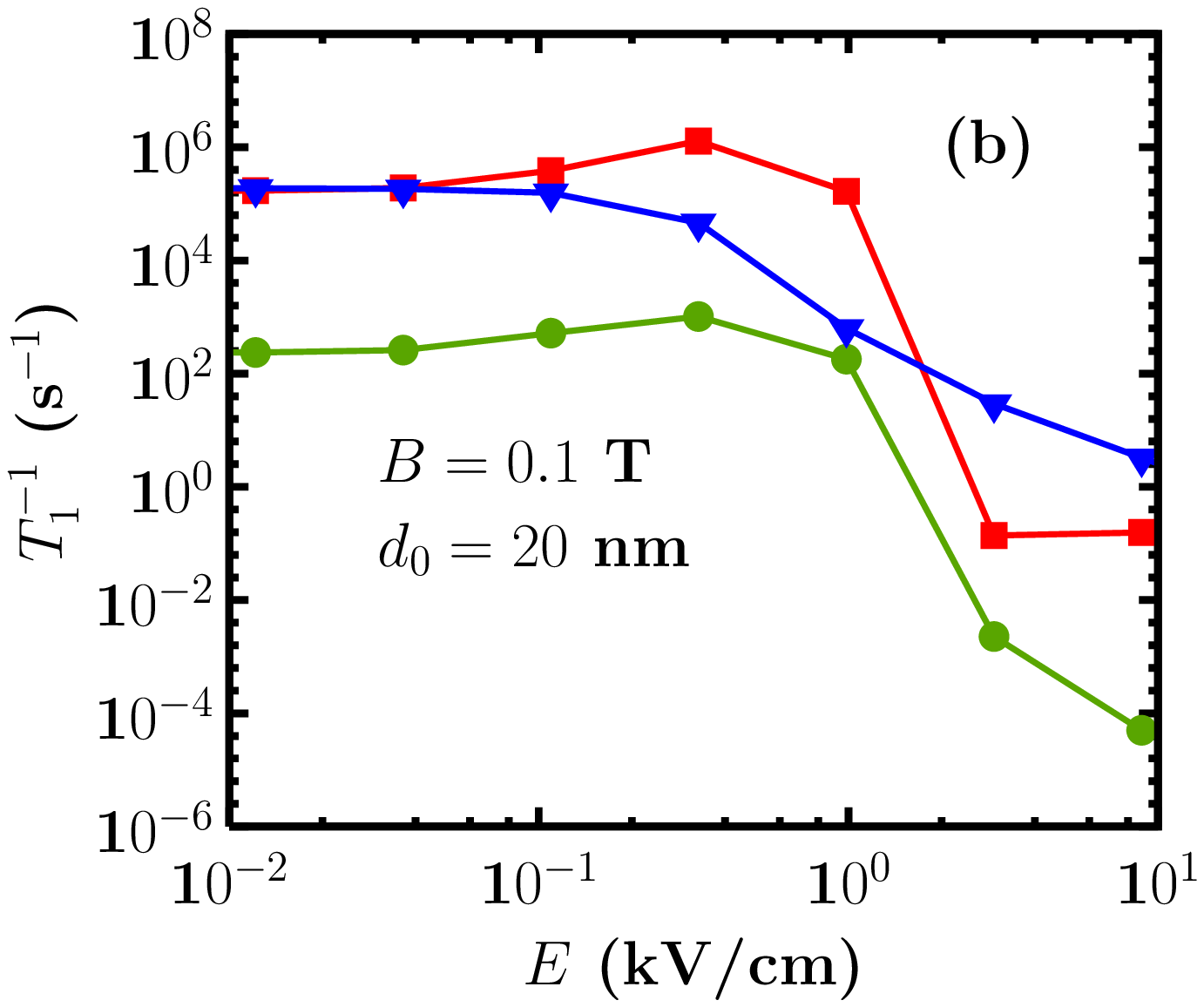}
    \includegraphics[height=4.2cm,width=5.1cm]{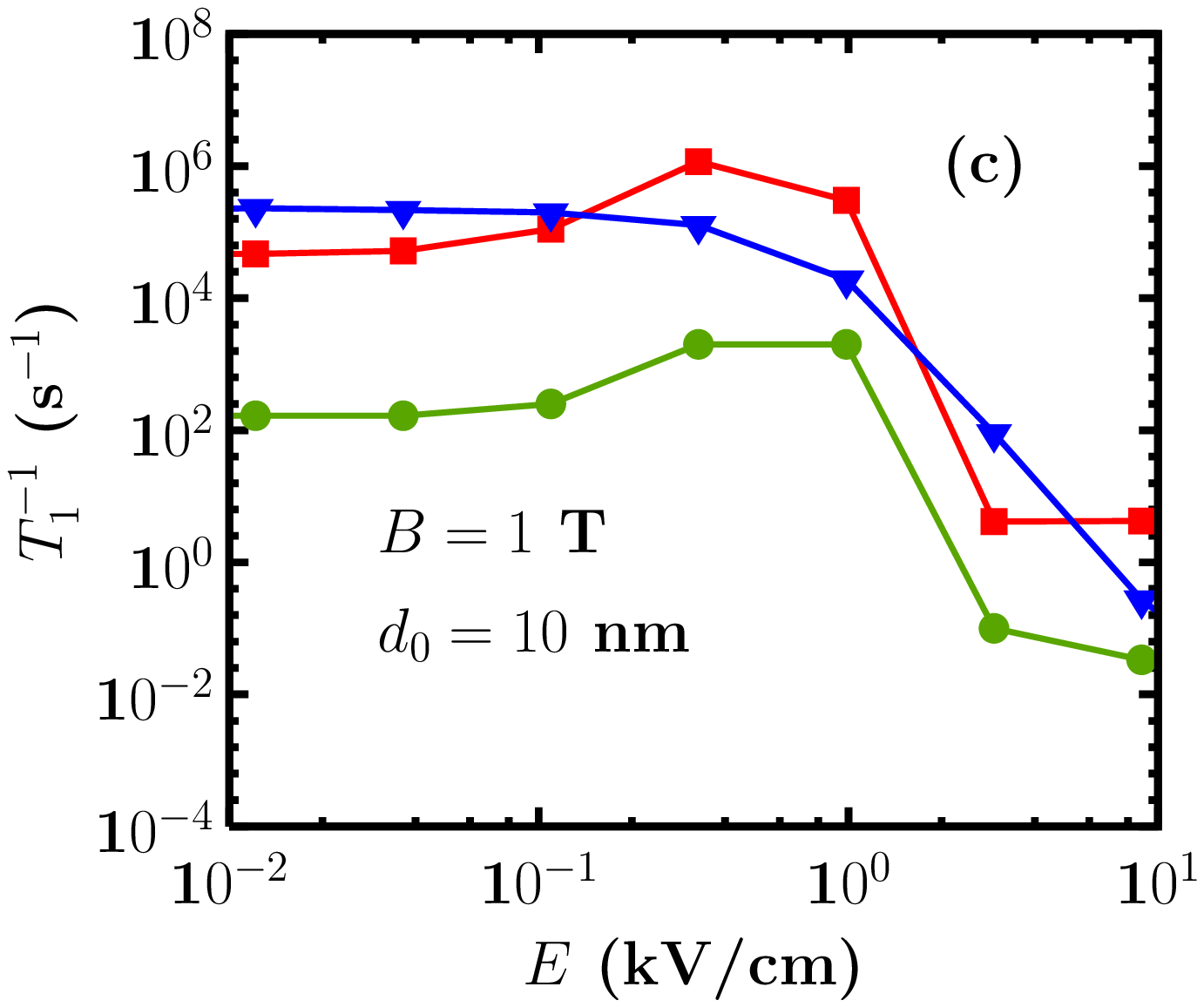}
    \includegraphics[height=4.2cm,width=5.1cm]{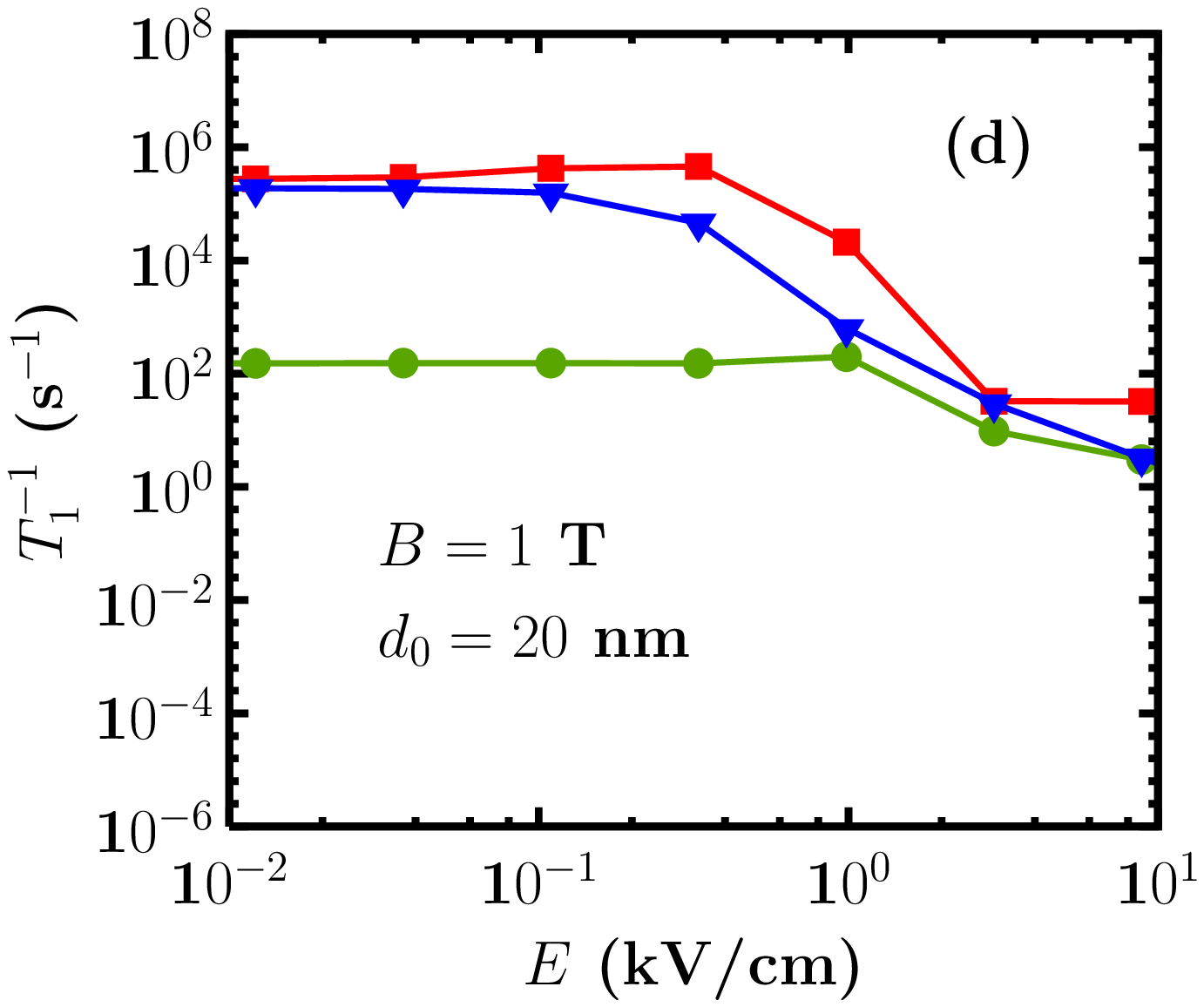}
\caption{(Color on line) $T_{1}^{-1}$ induced by
  different mechanisms
  {\em vs.} the electric field at different magnetic fields and dot diameters.
 $T=4$\ K. Curves with $\blacksquare$:
 by SOC together with the electron-BP scattering;
 Curves with  $\bullet$: by the
 direct spin-phonon coupling due to the phonon-induced strain; Curves
  with $\blacktriangledown$: by the  second-order process of the
  hyperfine interaction together with the BP ($V_{eI-ph}^{(3)}$).}
\label{fig2}
\end{figure}

\subsection{Spin relaxation time $T_{1}$  {\em vs.}  Electric field
  $E$}

The spin relaxation due to various mechanisms at
different magnetic fields $B$ and quantum dot diameters $d_{0}$
are plotted as functions of electric field $E$ in Fig.\ \ref{fig2}.
The temperature $T=4$\ K.
It is seen that with the increase of $E$, the spin relaxations induced
by all the three mechanisms almost keep unchanged for small $E$
(when $E<0.1$\ kV/cm), then increase a little and reach maximums
around $0.3$\ kV/cm. What is interesting is that when $E$ is
increased to around $1.1$\ kV/cm, the spin relaxations are suppressed
very quickly over a small window of $E$. Therefore, the total spin
relaxation can be controlled effectively with a small value of the
variation of the
bias field $\Delta E$.
 For example, in Fig.\
\ref{fig2}(a), the spin relaxation shows ten orders of magnitude
variation when the electric field $E$ changes from $0.1$ to
$10$\ kV/cm. This can be understood as following. All
the three mechanisms
are related to the electron-BP scattering which is affected
sensitively by the phonon wave length. The scattering becomes most efficient
when the phonon wave length is
comparable with the dot size. It is also known that the spin
relaxations between the
first and second subband are dominant for small electric field.\cite{Wang}
Therefore when $E<0.1$\ kV/cm, as the energy difference between the
lowest two subband along the
$z$-direction $\Delta {\cal E}_{12}$ is almost
constant (see Fig.\ \ref{fig1} (a)), the spin
relaxation  keeps nearly unchanged.
When the  phonon wave length is comparable with the dot size, the
 electron-phonon scattering becomes most efficient. Therefore the spin
relaxations show maximums. However, with the further increase of
energy difference $\Delta {\cal E}_{12}$ by the bias voltage, the phonon wave
length becomes larger than the dot size. Consequently the spin relaxation
decreases very quickly over a small window of $\Delta E$.

Now we focus on the variation magnitude of the spin relaxation
with the bias field under different conditions.
The largest variation of $T_{1}$
($10$ orders of magnitude) happens at small
magnetic field $B=0.1$\ T and small diameter $d_{0}=10$\ nm.
However, for larger $d_{0}$ ($20$\ nm in Fig.\ \ref{fig2}(b)) or larger
$B$ ($1$\ T in Fig.\ \ref{fig2}(c)), the
variations of the total spin relaxation decrease by several orders of
magnitude. It is further seen that when both $d_{0}$ and $B$ are
  increased [$d_{0}=20$\ nm and $B=1$\ T in Fig.\ \ref{fig2}(d)], the
  variations of the total spin relaxation is even smaller.
 This is because the spin relaxation induced
by the electron-BP interaction and
$V_{eI-ph}^{(3)}$ decreases with $B$ and $d_{0}$ in the
high electric field region, where electron is
mostly confined in one dot. This is similar to the single QD case.\cite{Jiang2}
Therefore to achieve large control of spin decoherence, the magnetic
field $B$ and dot diameter $d_{0}$ should be small.

\begin{figure}[bth]
  \centering
    \includegraphics[height=6cm,width=6cm]{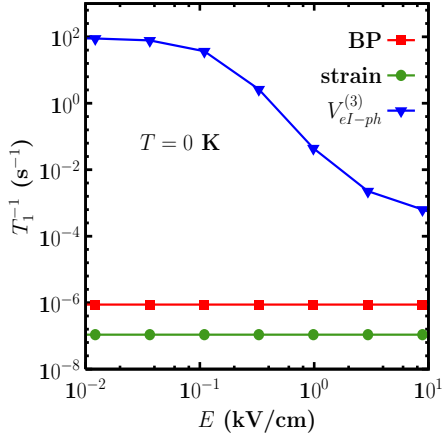}
\caption{(Color on line) $T_{1}^{-1}$  induced by
  different mechanisms
  {\em vs.} the electric field. In the calculation, $d_{0}=10$\ nm,
  $B=0.1$\ T and
  $T=0$\ K. Curve with $\blacksquare$:
 by SOC together with the electron-BP scattering; Curve
 with  $\bullet$:  by the
 direct spin-phonon coupling due to the phonon-induced strain; Curve
  with $\blacktriangledown$: by the
  second-order process of the
  hyperfine interaction together with the BP ($V_{eI-ph}^{(3)}$).}
\label{fig3}
\end{figure}

We further investigate the electric field dependence of spin
relaxation
at $T=0$\ K and the results are shown in Fig.\ \ref{fig3}.
It is interesting to see that
the spin relaxation induced by the second-order
process of hyperfine interaction combined with the electron-BP
scattering ($V_{eI-ph}^{(3)}$) still has large variation with $E$ ($5$
orders of magnitude variation when $E$ changes from $0.1$
to $10$\ kV/cm). However, the spin relaxations induced by the other two
mechanisms  keep almost unchanged.
This is because when $T=0$\ K, the electron
 only locates at the lowest orbital level and the spin
relaxation happens only
 between the lowest two Zeeman sublevels which
keeps unchanged  with
$E$. Therefore the large variations of spin relaxations induced by the
SOC together with the electron-BP scattering and the
strain-induced direct spin-phonon coupling no longer exist.
 However, for $V_{eI-ph}^{(3)}$, which is
the second-order process scattering, the middle states
$|m\rangle$ can be higher levels as the hyperfine interaction can
couple the spin-opposite states in different subband along the $z$-axis.
The energy differences between the middle states and the
initial/final states increase with $E$ and therefore the spin
relaxation induced by $V_{eI-ph}^{(3)}$ decreases with $E$
quickly.

\begin{figure}[bth]
  \centering
    \includegraphics[height=4.2cm,width=5.1cm]{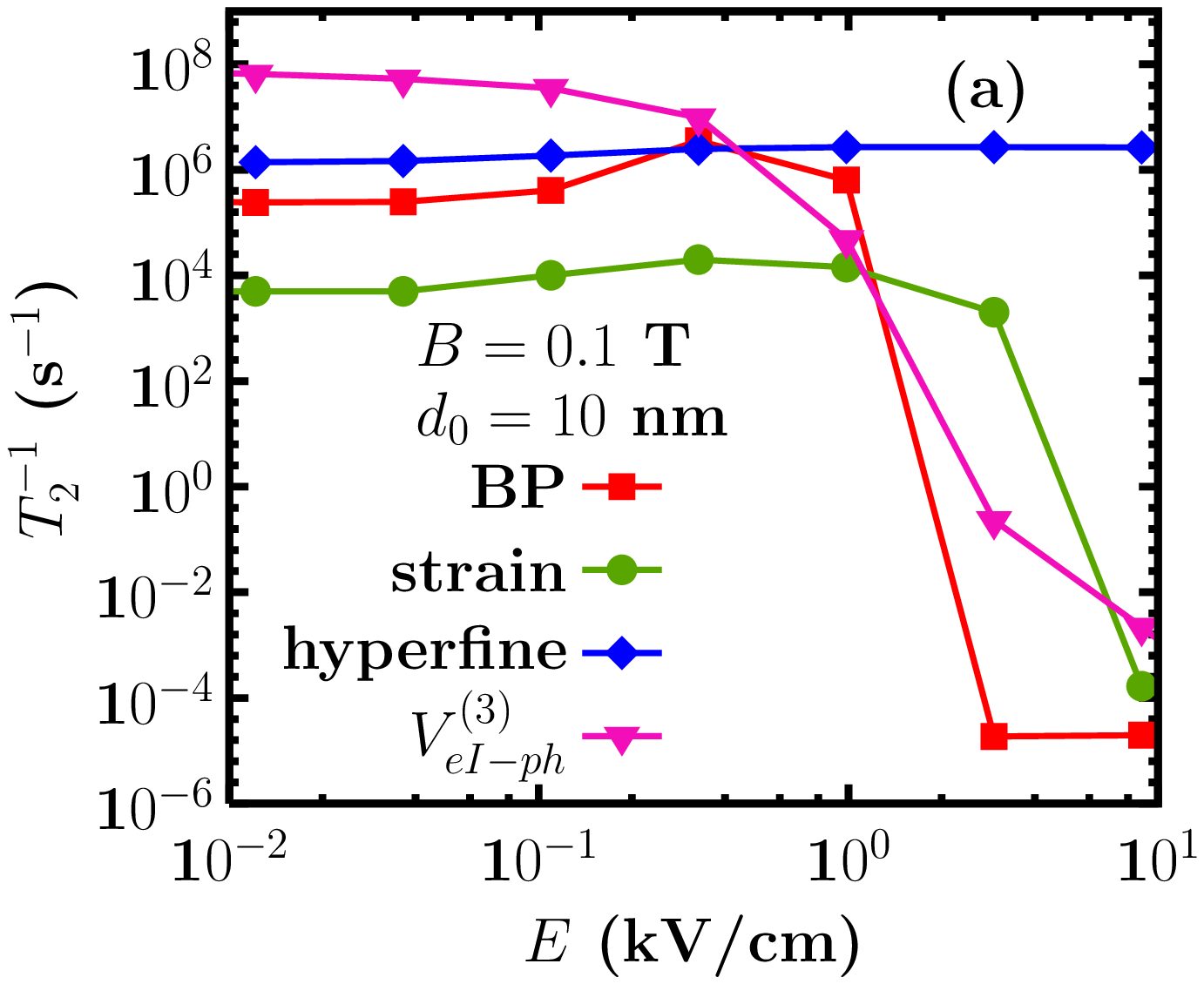}
    \includegraphics[height=4.2cm,width=5.1cm]{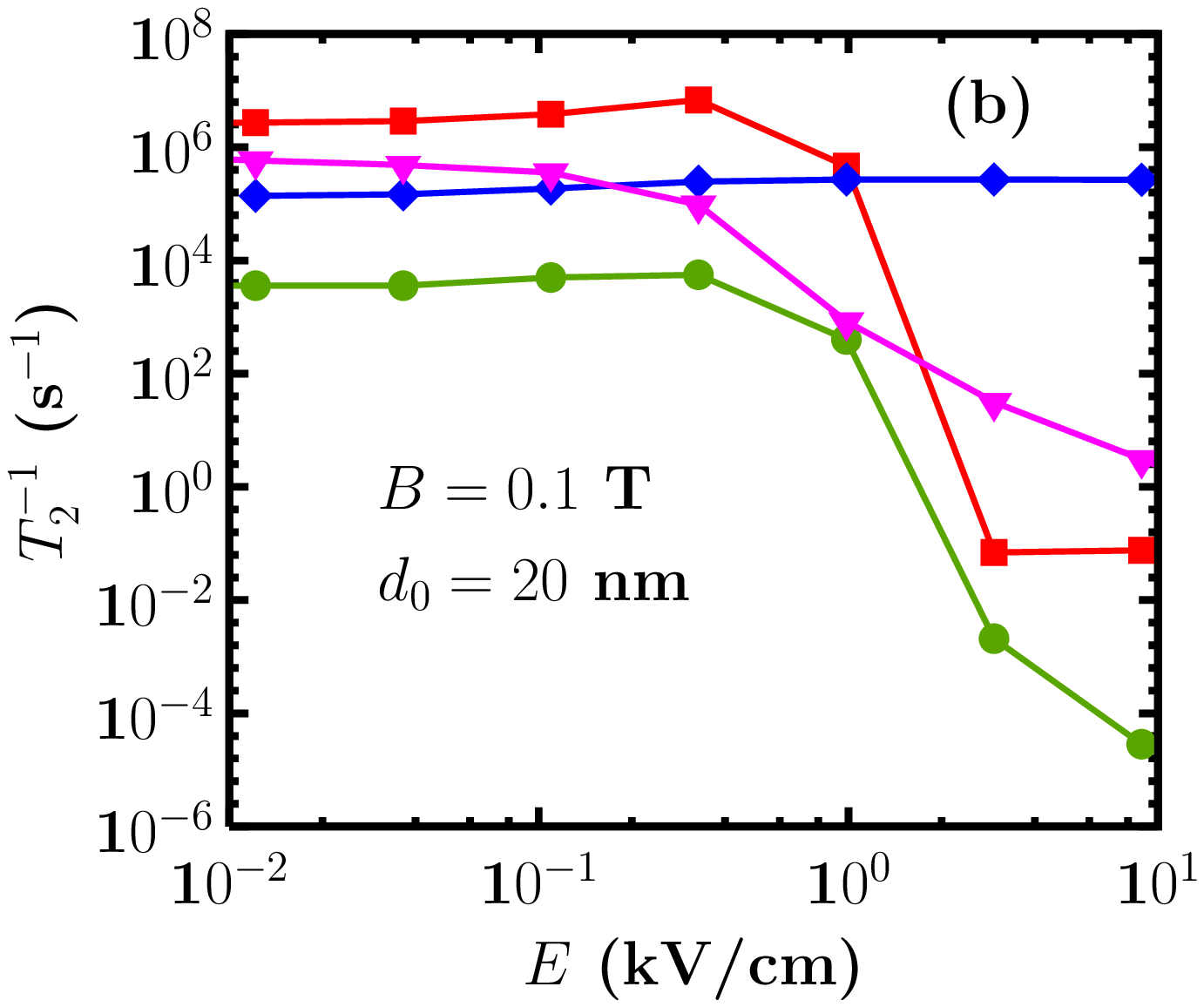}
    \includegraphics[height=4.2cm,width=5.1cm]{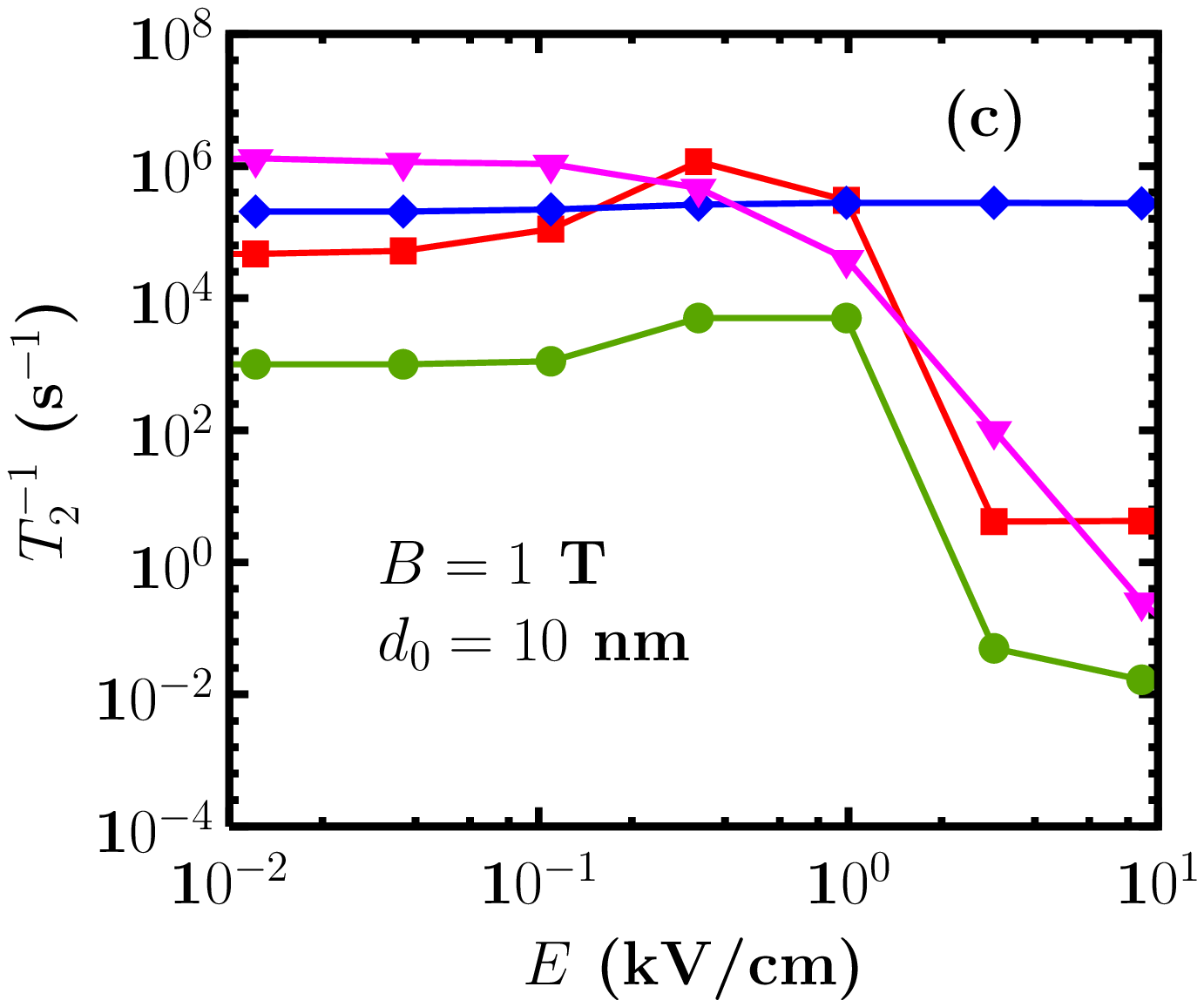}
    \includegraphics[height=4.2cm,width=5.1cm]{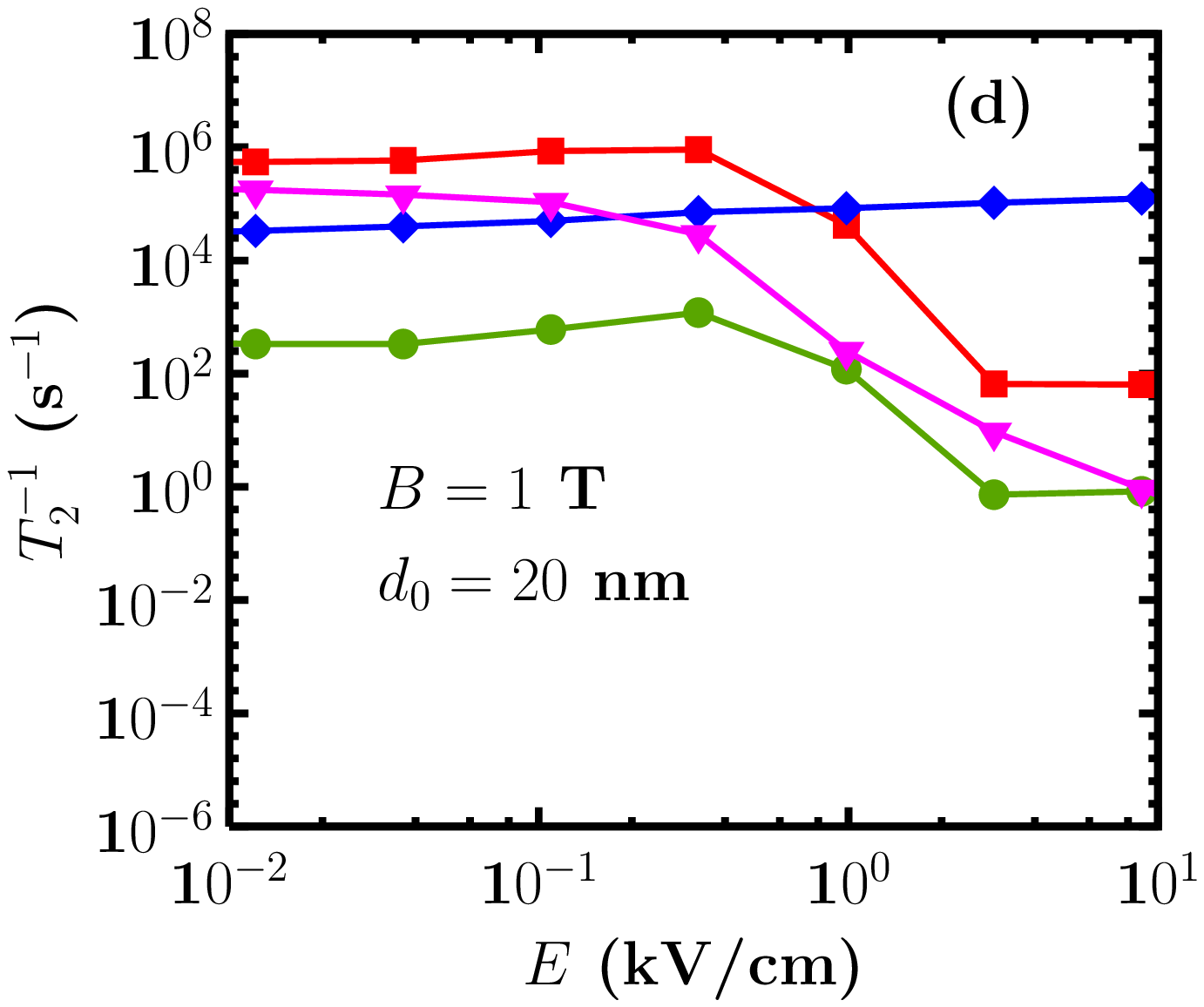}
\caption{(Color on line) $T_{1}^{-1}$ and $T_{2}^{-1}$ induced by
  different mechanisms
  {\em vs.} the electric field at different magnetic fields and dot diameters.
 $T=4$\ K. Curves with $\blacksquare$:
 by SOC together with the electron-BP scattering;
 Curves with  $\bullet$: by the
 direct spin-phonon coupling due to the phonon-induced strain; Curves
  with $\blacktriangledown$: by the  second-order process of the
  hyperfine interaction together with the BP ($V_{eI-ph}^{(3)}$);
  Curves with $\blacklozenge$:  by the
  hyperfine interaction.}
\label{fig4}
\end{figure}

\subsection{Spin dephasing time $T_{2}$  {\em vs.} Electric field $E$}

Now we turn to study the variation of spin dephasing with the electric
field $E$ and the results under different conditions are summarized in Fig.\
\ref{fig4}.
It is seen that the spin dephasing induced by the SOC together with
the electron-BP scattering, direct spin-phonon coupling due to the
phonon-induced strain and the second-order process of electron-BP
scattering combined with the hyperfine interaction always has several
orders of magnitude variation with $E$ at
different $d_{0}$ and $B$. However, the spin dephasing induced by the
hyperfine interaction only increases a little with $E$, which suppresses the
large variation of the spin dephasing induced by the other
three mechanisms. Nevertheless, there is still two orders of
magnitude variation of the spin dephasing when $B$ and $d_{0}$ are
small (Fig\ \ref{fig4}(a)).
The large variation of spin dephasing induced by the three mechanisms
related to electron-phonon scattering comes from the
fast increase of the energy difference $\Delta {\cal E}_{12}$,
similar to the analysis of spin relaxation. The spin
dephasing induced by the hyperfine interaction is not so sensitive
with $E$. This is because for the hyperfine interaction,
$T_{2}\approx E_{z} A^{-2} N$, which is obtained from the
pair-correlation approach,
with $E_{z}$, $A$ and $N$  being the Zeeman splitting energy, the hyperfine
interaction parameter and the nuclear number in the quantum
dot respectively.\cite{Yao}  With the increase of the electric field $E$, $E_{z}$ and $A$
keep unchanged and
 the wave function of the ground state
is gradually localized on one dot with lower potential.
This means the effective quantum dot size decreases and consequently
$N$ decreases. However, as the
effective quantum dot size decreases from two dots into one dot with
the bias field $E$, $N$ for large $E$ is only about a half of that for small $E$.
 Therefore, $T_{2}^{-1}$ for large field  is about two times of that for small field
   due to the decrease of $N$. However,
this increase of $T_{2}^{-1}$ induced by the hyperfine
interaction is
very small compared to that induced by other three mechanisms
(which have several orders of magnitude variation).
What should be pointed out is that
the variation of spin dephasing does not exist at $T=0$\ K. This is
because the hyperfine interaction is dominant for the spin
dephasing at $T=0$\ K, which keeps unchanged with $E$.

\section{Conclusion}

In conclusion, we propose a scheme to manipulate the
spin decoherence (both $T_1$ and $T_2$) in DQD system
by a small gate voltage.
 Up to ten orders of magnitude of spin relaxation and up to two
 orders of magnitude of spin dephasing can be
 obtained. To obtain
large variation of spin decoherence, the inter-dot distance and/or
the barrier hight should be large enough in order to guarantee that the
two QDs are nearly
independent for the small bias voltage.
At the same time, the effective diameter and magnetic field should
be small to get as large  variation as possible. Finally, based on
the present available experimental technology, the DQD system
applied in this paper can be realized easily.\cite{Ono,Austing}

\begin{acknowledgments}

This work was supported by the Natural Science Foundation of China
under Grant Nos.\ 10574120 and 10725417,
the National Basic Research Program of
China under Grant No.\ 2006CB922005 and
the Innovation Project of Chinese Academy of Sciences.

\end{acknowledgments}

\end{document}